\documentclass[12pt,english]{elsarticle}
\usepackage[T1]{fontenc}
\usepackage[latin9]{inputenc}
\usepackage{mathrsfs}
\usepackage{amsmath}
\usepackage{amssymb}

\makeatletter
 \newenvironment{proofQED}{\begin{proof}}{\qed\end{proof}}

\RequirePackage{fix-cm}

\usepackage{todonotes}

\usepackage{transparent}

\newtheorem{theorem}{Theorem}\newtheorem{definition}{Definition}\newtheorem{lemma}{Lemma}\newdefinition{remark}{Remark}
\newdefinition{example}{Example}
\newdefinition{corollary}{Corollary}
\newproof{proof}{Proof}
\newproof{thesis}{Thesis}

\usepackage{tikz}

\usepackage{babel}

\makeatother

\usepackage{babel}
\begin{document}

\title{When is the condition of order preservation met?}

\author{Konrad Ku\l akowski}

\ead{konrad.kulakowski@agh.edu.pl}

\address{AGH University of Science and Technology, Poland}

\author{Ji\v{r}{\'\i} Mazurek}

\ead{mazurek@opf.slu.cz}

\address{Silesian University Opava, Czech Republic}

\author{Jaroslav Ram{\'\i}k}

\ead{ramik@opf.slu.cz}

\address{Silesian University Opava, Czech Republic}

\author{Michael Soltys}

\ead{michael.soltys@csuci.edu}

\address{California State University Channel Islands, USA}
\begin{abstract}
This article explores a relationship between inconsistency in the
pairwise comparisons method and conditions of order preservation. A
pairwise comparisons matrix with elements from an alo-group is
investigated. This approach allows for a generalization of previous
results. Sufficient conditions for order preservation based on the
properties of elements of pairwise comparisons matrix are derived. A
numerical example is presented.
\end{abstract}
\begin{keyword}
pairwise comparisons\sep alo-groups \sep EVM\sep GMM \sep COP \sep AHP 
\end{keyword}
\maketitle

\section{Introduction }

The first documented use of comparisons by pairs dates back to the
XIII century \cite{Colomer2011rlfa}. Later, the method was developed
by \emph{Fechner} \cite{Fechner1966eop}, \emph{Thurstone}
\cite{Thurstone27aloc} and \emph{Saaty} \cite{Saaty1977asmf}.  Saaty
proposed the seminal \emph{Analytic Hierarchy Process} \emph{(AHP)}
extension to pairwise comparisons (henceforth abbreviated as\emph{
PC}) theory, which is a framework for dealing with a large number of
criteria. At the beginning of the XX century the method was used in
psychometrics and psychophysics \cite{Thurstone27aloc}. Now it is
considered part of decision theory \cite{Saaty2005taha}.  Its utility
has been confirmed in numerous examples
\cite{Vaidya2006ahpa,Ho2008iahp,Liberatore2008tahp,Subramanian2012aroa}.
Despite the relative maturity of the area, it still invites further
exploration. Examples of new exploration are the \emph{Rough Set}
approach \cite{Greco2011fk}, voting systems
\cite{Faliszewski2009lacv}, fuzzy \emph{PC} relation handling
\cite{Mikhailov2003dpff,Yuen2013fcnp,Ramik2015ibfp}, incomplete
\emph{PC} relation \cite{Bozoki2010ooco,Fedrizzi2007ipca},
non-numerical rankings \cite{Janicki2012oapc,Janicki2011ropc}, nonreciprocal \emph{PC}
relation properties \cite{Fulop2012oscp}, rankings with the reference
set of alternatives \cite{Kulakowski2013hrea,kulakowski2014hreg},
applications to software correctness and cybersecurity
\cite{soltys-fund-2015} and others. Further references:
\cite{Smith2004aada,Ishizaka2009ahpa,Kou2016pcmi}.

Although the \emph{AHP} is a popular method for multiple-criteria
decision making, it is often criticized as in \cite{Brunelli2015itta}.
An important objection to \emph{AHP} originates from \emph{Bana e
Costa} and \emph{Vansnick}, \cite{BanaeCosta2008acao}, where the
authors
formulated a so called \emph{COP} (conditions of order
preservations) and proved that the priority deriving method followed
in \emph{AHP} does not meet \emph{COP}. This phenomenon is not,
however,
an inherent problem of \emph{AHP}
\cite{Kulakowski2015noop,Ho2017tsot}.  Instead it is the result of
inconsistency and the size of the differences between the
alternatives. Further study of \emph{COP}, the {\em Eigenvalue Method
(EVM)} and inconsistency can
be found in \cite{Kulakowski2015otpo}. In particular the work brings a
theorem showing dependency of \emph{COP} and \emph{Koczkodaj}
inconsistency index \cite{Koczkodaj1993ando} in the context of
\emph{EVM}.

Verifying whether for a certain set of pairwise comparisons \emph{COP}
is met requires ranking calculation. As \emph{COP} was originally
formulated in the context of \emph{EVM} the criticism
caused by \cite{BanaeCosta2008acao} was directed at \emph{EVM} and
\emph{AHP}. An attentive reader will notice, however, that
neither \emph{COP} as such, nor the notion of consistency understood
as the cardinal transitivity \cite[p. 158]{Bozoki2008osak},
depend on prioritization methods. Hence the question arises whether
the relationship between \emph{COP} and inconsistency is of a general
nature and if so, how does this relationship look like?

The present work is an attempt to answer this vital question. In order
to emphasize the general nature of the relationship between COP and
inconsistency we decided to use pairwise comparisons based on ordered
abelian groups. These general results are presented in
Section~\ref{sec:Condition-of-Order}. Although \emph{AHP} was defined
in the context of \emph{EVM} other methods of prioritization are
becoming more and more popular. The \emph{Geometric Mean Method (GMM)}
may serve as an example \cite{Ishizaka2011rotm}. Therefore, in
Section~\ref{sec:Generalized-Geometric-Mean} we redefine
\emph{GMM} in the context of alo-groups and show that meeting the
\emph{COP} criteria under this generalized \emph{GMM} depends on the
locally defined inconsistency. Similarly to
\cite{Kulakowski2015otpo} we adopt the 
inconsistency index \cite{Koczkodaj1993ando}. 

\section{Preliminaries, pairwise comparisons method}

The input data for the \emph{PC} method is a \emph{PC} matrix
$C=[c_{ij}]$, $i,j\in\{1,\ldots,n\}$, that expresses a weight function
$R$ with domain the finite set of alternatives
$A\overset{\textit{}}{=}\{a_{i}\in\mathscr{A}|i\in\{1,\ldots,n\}\}$.
The set $\mathscr{A}$ is a non empty universe of alternatives and
$R(a_{i},a_{j})=c_{ij}$. The values of comparisons $c_{ij}$ indicate
the relative importance of alternatives $a_{i}$ with respect to
$a_{j}$.  Here, the elements $c_{ij}$ of \emph{PC} matrix $C=[c_{ij}]$
belong to $\mathcal{G}$, an alo-group which will be defined bellow. 

An \emph{abelian group} is a set, $G$, together with an operation
$\odot$ (read ``{\em odot}'') that combines any two elements
$a,b\in G$ to form another element in $G$ denoted by $a\odot b$,
see \cite{Bourbaki1990aii,CavalloDApuzzo2009aguf}. The symbol $\odot$
is a general placeholder for some concretely given operation. $(G,\odot)$
satisfies the following requirements known as the \emph{abelian group
axioms}, particularly: \emph{commutativity}, \emph{associativity},
there exists an \emph{identity element} $e\in G$ and for each element
$a\in G$ there exists an element $a^{(-1)}\in G$ called the \emph{inverse
element to $a$}.

The \emph{inverse operation} \index{inverse operation} $\div$ to
$\odot$ is defined for all $a,b\in G$ as follows:
\begin{equation}
a\div b=a\odot b^{(-1)}.\label{1}
\end{equation}
Note that the inverse operation is not necessarily associative.

An ordered triple $(G,\odot,\leq)$ is said to be an \emph{abelian
linearly ordered group}, \emph{alo-group} for short, if $(G,\odot)$ is
a group, $\leq$ is a linear order on $G$, and for all $a,b,c\in G$:
\begin{equation}
a\leq b\textrm{ implies }a\odot c\leq b\odot c.\label{2},
\end{equation}
in other words, $\odot$ respects $\le$.

If $\mathcal{G}=(G,\odot,\leq)$ is an alo-group, then $G$ is naturally
equipped with the order topology induced by $\leq$ and $G\times G$
is equipped with the related product topology. We say that $\mathcal{G}$
is a \emph{continuous alo-group} if $\odot$ is continuous on $G\times G$.

By definition, an alo-group $\mathcal{G}$ is a lattice ordered group.
Hence, there exists $\max\{a,b\}$, for each pair $(a,b)\in G\times G$
. Nevertheless, a nontrivial alo-group $\mathcal{G}=(G,\odot,\leq)$
has neither the greatest element nor the least element.

Because of the associative property, the operation $\odot$ can be
extended by induction to $n$-ary operation.

$\mathcal{G}=(G,\odot,\leq)$ is \emph{divisible} if for each positive
integer $n$ and each $a\in G$ there exists the ($n$)-th root of
$a$ denoted by $a^{(1/n)}$, i.e., $\left(a^{(1/n)}\right)^{(n)}=a$.

Let \ $\mathcal{G}=(G,\odot,\leq)$ be an alo-group. Then the function
$\left\Vert .\right\Vert :G\rightarrow G$ defined for each $a$ $\in$
$G$ by 
\begin{equation}
\|a\|=\max\{a,a^{(-1)}\}\label{3}
\end{equation}
is called a $\mathcal{G}$-\emph{norm}.

The operation $d:G\times G\rightarrow G$ defined by $d(a,b)=\left\Vert a\div b\right\Vert $
for all $a,b$ $\in$ $G$ is called a $\mathcal{G}$-\emph{distance}.
Next, we present the well known examples of alo-groups; for more details
see also \cite{CavalloDApuzzo2009aguf}, or, \cite{Ramik2015ibfp}. 
\begin{example}
\label{E1} \emph{Additive alo-group} $\mathcal{R}=(\mathbb{R},+,\leq)$
is a continuous alo-group with:
$e=0,\ a^{(-1)}=-a,\ a^{(n)}=n\cdot a$. 

\label{E2} \emph{Multiplicative alo-group} $\mathcal{R}_{+}=(\mathbb{R}_{+},\cdot,\leq)$
is a continuous alo-group with:
 $e=1,\ a^{(-1)}=a^{-1}=1/a,\ a^{(n)}=a^{n}.$ Here, by `$\cdot$' we
denote the usual operation of multiplication. 

\label{E3} \emph{Fuzzy additive alo-group} $\mathcal{R}_{a}$=$(\mathbb{R},+_{f},\leq)$,
see \cite{Ramik2015ibfp}, is a continuous alo-group with: 
\[
a+_{f}b=a+b-0.5,\ e=0.5,\ a^{(-1)}=1-a,a^{(n)}=n\cdot a-\frac{n-1}{2}.
\]

\label{E4} \emph{Fuzzy multiplicative alo-group} ${\bf {]0,1[_{m}}}$=$(]0,1[,\bullet_{f},\leq)$,
see \cite{CavalloDApuzzo2009aguf}, is a continuous alo-group with:
\[
a\bullet_{f}b=\frac{ab}{ab+(1-a)(1-b)},e=0.5,a^{(-1)}=1-a.
\]
\end{example}

\begin{remark}
Usually, the \emph{PC} method is used with a multiplicative \emph{PC}
matrix, i.e., with multiplicative alo-group, see e.g.
\cite{Saaty1977asmf,Kulakowski2015noop}.  Then the relative importance
of an alternative is multiplied with the relative importance of the
other alternatives when considering a chain of alternatives. Now, our
approach based on a more general concept applying alo-groups enables
to extend the properties of the multiplicative system to the whole
class of pairwise comparisons systems.  The four instances listed in
Example 1 show some useful non-trivial cases. 
\end{remark}

Now, we define two important properties of a \emph{PC} matrix. 
\begin{definition}
\label{def:A-matrix-recip}A \emph{PC} matrix $C=[c_{ij}],c_{ij}\in G$,
is said to be $\odot$-\emph{reciprocal} if 
\begin{equation}
c_{ij}\odot c_{ji}=e\textrm{ for all }i,j\in\{1,\ldots,n\},\label{4}
\end{equation}
or, equivalently, 
\begin{equation}
c_{ji}=c_{ij}^{(-1)}\textrm{ for all }i,j\in\{1,\ldots,n\},\label{5}
\end{equation}
and it is said to be $\odot$-\emph{consistent} if 
\begin{equation}
c_{ij}\odot c_{jk}\odot c_{ki}=e\textrm{ for all }i,j\in\{1,\ldots,n\}.\label{6}
\end{equation}
\end{definition}

Evidently, if $C$ is $\odot$-consistent, then it is also
$\odot$-reciprocal, but not vice versa. Since the \emph{PC} matrix
usually contains subjective evaluations provided by (human) experts,
the information contained therein may be $\odot$-\emph{inconsistent}.
That is, a triad of values $c_{ij},c_{jk},c_{ki}$ in $C$ may exist for
which $c_{ij}\odot c_{jk}\odot c_{ki}\neq e$.  In other words,
different ways of estimating the value of a pair of alternatives may
lead to different results. This fact leads to the concept of an
$\odot$-inconsistency index describing the extent to which the matrix
$C$ is $\odot$-inconsistent.

\section{Priority deriving methods}

There are a number of inconsistency indexes associated with deriving
PC rankings, including the \emph{Eigenvector Method}
\cite{Saaty1977asmf}, the \emph{Least Squares Method}, the 
\emph{Chi Squares
Method} \cite{Bozoki2008osak}, \emph{Koczkodaj's} \emph{distance based
inconsistency index}\footnote{\emph{An alternate form of this
definition can be found in \cite{Koczkodaj2014oaoi}.}}
\cite{Koczkodaj1993ando}, the \emph{Geometric Mean Method (GMM)} and
others. The three most prominent methods are described below.

The result of the pairwise comparisons method is a ranking---a mapping
that assigns values to the concepts. Formally, it can be defined as
the following function. 
\begin{definition}
\label{def:The-ranking-function}The \emph{ranking function for} $A$
(the \emph{ranking of} $A$) is a function $w:A\rightarrow\mathbb{R}_{+}$
that assigns to every alternative from $A\subset\mathscr{A}$ a positive
value from $\mathbb{R}_{+}$. 
\end{definition}

In other words, $w(a)$ represents the ranking value for $a\in A$.
The $w$ function is usually written in the form of a vector of weights,
i.e., $w\overset{\textit{df}}{=}\left[w(a_{1}),\ldots w(a_{n})\right]^{T}$
and is called the \emph{priority vector}.

Now, for the moment, we consider the usual multiplicative alo-group
$\mathcal{R}_{+}=(\mathbb{R}_{+},\cdot,\leq)$.  The eigenvalue based
consistency index $CI(C)$ called \emph{Saaty's index} of $n\times n$
reciprocal matrix $C=[c_{ij}]$ is defined as: 
\begin{equation}
\textit{CI(C)}=\frac{\lambda_{\textit{max}}-n}{n-1},\label{7}
\end{equation}
where $\lambda_{\textit{max}}$ is the principal eigenvalue of $C$.

The value $\lambda_{\textit{max}}\geq n$ and $\lambda_{\textit{max}}=n$
only if $C$ is consistent \cite{Saaty2013dk}. Vector $w$ is determined
as the rescaled principal eigenvector of $C$. Thus, assuming that
$Cw_{max}=\lambda_{\textit{max}}w_{max}$ the priority vector $w$
is 
\[
w=\gamma\left[w_{\textit{max}}(a_{1}),\ldots,w_{\textit{max}}(a_{n})\right]^{T},
\]
where $\gamma$ is a scaling factor. Usually it is assumed that $\gamma=\left(\sum_{i=1}^{n}w_{\textit{max}}(a_{i})\right)^{-1}$.
This method is called the \emph{Eigenvector Method (EVM)}.

Here, we consider a space of evaluations; an alo-group with the only
one binary operation, particularly, $\odot=\cdot$. Therefore, the
Saaty's index cannot be defined, because two group operations in (\ref{7}),
e.g. $\cdot$, +, are necessary. In what follows, we shall not deal
with the Eigenvector Method.

\emph{Koczkodaj's inconsistency index} $KI$ of $n\times n$ and ($n>2)$
reciprocal matrix $C=[c_{ij}]$ is defined as 
\begin{equation}
\textit{KI(C)}=\underset{i,j,k\in\{1,\ldots,n\}}{\max}\left\{ 1-\min\left\{ \frac{c_{ij}}{c_{ik}c_{kj}},\frac{c_{ik}c_{kj}}{c_{ij}}\right\} \right\} .\label{8}
\end{equation}

Similarly, as we consider here a space alo-group with one binary
operation, particularly, $\odot=\cdot$, the \emph{Koczkodaj's} inconsistency
index cannot be defined, as two group operations (\ref{8}), e.g.
$\cdot$, +/-, are necessary. That is why we shall not deal with the
\emph{Koczkodaj's} inconsistency index. Later on (Theorem \ref{theorem3}),
however, we shall derive a relationship between \emph{Koczkodaj's}
inconsistency index and the generalized inconsistency index (which
will be also defined later) in the multiplicative alo-group $\mathcal{R}_{+}=(\mathbb{R}_{+},\cdot,\leq)$
together with the additional field operation +, see Example \ref{ex:example}.

One of the most important, and still gaining in importance, methods
of deriving priorities from pairwise comparisons has been proposed
by \emph{Crawford} \cite{Crawford1987tgmp}. According to this approach,
referred in the literature as \emph{geometric mean method (GMM)} the
weight of $i$-th alternative is given by the geometric mean of the $i$-th
row of $C$. Thus, the priority vector is given as 
\begin{equation}
w=\gamma\left[\left(\prod_{r=1}^{n}c_{1r}\right)^{\frac{1}{n}},\ldots,\left(\prod_{r=1}^{n}c_{nr}\right)^{\frac{1}{n}}\right]^{T},\label{9}
\end{equation}
where $\gamma$ is a scaling factor. As previously,
$\gamma=\left(\sum_{i=1}^{n}w_{\max}(a_{i})\right)^{-1}$.
Following \cite{Saaty1977asmf}, we obtain the following definition.
\begin{definition}
\label{def:The global error index} Let $C=[c_{ij}]$ be a reciprocal
PC matrix. For each pair $i,j\in\{1,\ldots,n\}$, and a priority vector
$w=\left[w(a_{1}),\ldots,w(a_{n})\right]^{T}$, a \emph{local error
index} $\epsilon(i,j,w)$ is given as
\begin{equation}
\epsilon(i,j,w)\overset{\textit{df}}{=}c_{ij}\odot w(a_{j})\div w(a_{i}),\label{10}
\end{equation}
and similarly (as in \cite{Kulakowski2015noop}) let us define 
\begin{equation}
\begin{array}{rl}
\mathscr{E}(i,j,w) & \overset{\textit{df}}{=}\max\{\epsilon(i,j,w),(\epsilon(i,j,w))^{(-1)}\}.\label{11}\end{array}
\end{equation}
The \emph{global error index} $\mathscr{E}(C,w)$ for the PC matrix
$C$ and a priority vector $w=(w_{1},\ldots,w_{n})$, is the maximal
value of $\mathscr{E}(i,j,w)$, i.e., 
\begin{equation}
\mathscr{E}(C,w)\overset{\textit{df}}{=}\max_{i,j\in\{1,\ldots,n\}}\mathscr{E}(i,j,w).\label{12}
\end{equation}
\end{definition}

Now, let us derive the following properties of $\mathscr{E}(C,w)$.
\begin{lemma}
\label{Lemma1} Let $C=[c_{ij}]$ be a reciprocal PC matrix and $w=\left[w(a_{1}),\ldots,w(a_{n})\right]^{T}$
be a priority vector. Then 
\begin{equation}
\mathscr{E}(C,w)\geq e,\label{15}
\end{equation}
moreover, if 
\[
\mathscr{E}(C,w))=e,
\]
then $C$ is $\odot$-consistent. 
\end{lemma}

\begin{proofQED}
Either $\epsilon(i,j,w)\geq e$, or, $\epsilon(i,j,w)\leq e$, then
$\epsilon(i,j,w))^{(-1)}\geq e.$ Hence, 
\[
\mathscr{E}(i,j,w)=\max\{\epsilon(i,j,w),(\epsilon(i,j,w))^{(-1)}\}\geq e.
\]
By (\ref{12}) we obtain 
\[
\mathscr{E}(C,w)\geq e.
\]
Moreover, let $\mathscr{E}(C,w)=e.$ Then by (\ref{11}), (\ref{12})
for all $i,j\in\{1,\ldots,n\}$, it holds 
\[
\epsilon(i,j,w)=c_{ij}\odot w(a_{j})\div w(a_{i})=e,
\]
hence, equivalently, 
\[
c_{ij}=w(a_{i})\div w(a_{j}).
\]
Then we obtain 
\[
c_{ij}\odot c_{jk}\odot c_{ki}=w(a_{i})\div w(a_{j})\odot w(a_{j})\div w(a_{k})\odot w(a_{k})\div w(a_{i})=e,
\]
hence by (\ref{6}), $C$ is $\odot$-consistent. 
\end{proofQED}

\begin{remark}
The global error index $\mathscr{E}(C,w)$ depends not only on the
elements $c_{ij}$ of \emph{PC} matrix $C$, but also on the priority
vector $w$. It is, however, always greater or equal to the identity
element $e\in G$. If the global error index of $C$ is equal to $e$
then \emph{PC} matrix $C$ is $\odot$-consistent. Later, in
Lemma~\ref{lemma2}, we will show that if \emph{PC} matrix $C$ is
$\odot$-consistent, then there exists a priority vector $w$ such that
$\mathscr{E}(C,w)=e$ holds. 
\end{remark}

\section{Condition of Order Preservation\label{sec:Condition-of-Order}}

In \cite{BanaeCosta2008acao} \emph{Bana e} \emph{Costa} and \emph{Vansnick}
formulate two conditions of order preservations. Here, we formulate
these conditions in a more general setting, i.e., for alo-groups. The
first, \emph{the preservation of order preference condition} \emph{(POP}),
claims that the ranking result in relation to the given pair of alternatives
$(a_{i},a_{j})$ should not break with the expert judgment, that is,
if for a pair of alternatives $a_{i},a_{j}\in\mathscr{A}$ such that
$a_{i}$ dominates $a_{j}$ ($c_{ij}>e$) then:
\begin{equation}
w(a_{i})>w(a_{j}),\text{ or, equivalently }w(a_{i})\div w(a_{j})>e.\label{17}
\end{equation}
Here, $w(a_{k}),k=1,2,...,n$, are individual weights of a priority
vector $w$.

The second one \emph{the preservation of order of intensity of preference
condition }(\emph{POIP),} claims that 
if $a_{i}$ dominates $a_{j}$ more than $a_{k}$ dominates $a_{l}$
($a_{i},a_{j},a_{k},a_{l}\in\mathscr{A}$), i.e.,
if $c_{ij}>e$, $c_{kl}>e$ and $c_{ij}>c_{kl}$ then 
\begin{equation}
w(a_{i})\div w(a_{j})>w(a_{k})\div w(a_{l}).\label{18}
\end{equation}
We show that POP and POIP condition is satisfied if the PC matrix
is $\odot$-consistent. We start with the well known necessary and
sufficient condition for a PC matrix to be $\odot$-consistent, see
also \cite{Ramik2015ibfp}.
\begin{lemma}
\label{lemma2} Let $C=[c_{ij}]$ be an $\odot$-reciprocal PC matrix.
Then $C$ is $\odot$-consistent if and only if there exists a priority
vector $w=\left[w(a_{1}),\ldots,w(a_{n})\right]^{T}$ such that for
all $i,j\in\{1,\ldots,n\}$ 
\begin{equation}
w(a_{i})\div w(a_{j})=c_{ij}.\label{19}
\end{equation}
\end{lemma}

\begin{proofQED}
Suppose that $C=[c_{ij}]$ is $\odot$-consistent, then by Definition
\ref{def:A-matrix-recip} 
\[
c_{ij}\odot c_{jk}\odot c_{ki}=e\text{ for all }i,j,k\in\{1,\ldots,n\},
\]
or, equivalently, 
\[
c_{ij}\odot c_{jk}=c_{ik}.
\]
Let $w=(w_{1},\ldots,w_{n})$ be given by 
\begin{equation}
w(a_{i})=\delta \odot \left(\bigodot_{r=1}^{n}c_{ir}\right)^{(\frac{1}{n})},i\in\{1,2,...,n\},\label{140}
\end{equation}
where $\delta$ is a scaling factor, $\delta=\left(\bigodot_{r=1}^{n}c_{1r}\right)^{(-\frac{1}{n})}\odot\ldots\odot\left(\bigodot_{r=1}^{n}c_{nr}\right)^{(-\frac{1}{n})}.$
 Then we obtain by consistency condition $c_{ir}\odot c_{rj}=c_{ij}$
\[
w(a_{i})\div w(a_{j})=\delta \odot \left(\bigodot_{r=1}^{n}c_{ir}\right)^{(\frac{1}{n})}\odot\left(\delta\bigodot_{r=1}^{n}c_{jr}\right)^{(-\frac{1}{n})}=
\]
\[
=\left(\bigodot_{r=1}^{n}(c_{ir}\odot c_{rj})\right)^{(\frac{1}{n})}=\left(c_{ij}^{(n)}\right)^{(\frac{1}{n})}=c_{ij},
\]
hence, (\ref{19}) is satisfied.
On the other hand, let condition (\ref{19}) be satisfied. Then for
each $i,j,k\in\{1,\ldots,n\}$ we obtain 
\[
c_{ij}\odot c_{jk}\odot c_{ki}=(w(a_{i})\div w(a_{j}))\odot(w(a_{j})\div w(a_{k}))\odot(w(a_{k})\div w(a_{i}))=
\]
\[
=w(a_{i})\odot w(a_{j})^{(-1)}\odot w(a_{j})\odot w(a_{k})^{(-1)}\odot w(a_{k})\odot w(a_{i})^{(-1)}=e,
\]
hence, $C$ is $\odot$-consistent. 
\end{proofQED}

\begin{theorem}
\label{theorem1} Let $C=[c_{ij}]$ be an $\odot$-reciprocal PC matrix,
and let $w=\left[w(a_{1}),\ldots,w(a_{n})\right]^{T}$ be a priority vector,
let $i,j,k,l\in\{1,\ldots,n\}$.
 If $C$ is $\odot$-consistent then condition POP is satisfied, i.e.
$c_{ij}>e$ implies $w_{i}>w_{j}$. Moreover, if $c_{ij}>c_{kl}$,
then condition POIP is also satisfied, i.e., $c_{ij}>c_{kl}$ implies
$w_{i}\div w_{j}>w_{k}\div w_{l}$. 
\end{theorem}

\begin{proofQED}
Suppose that $C=[c_{ij}]$ is $\odot$-consistent. If for some $i,j\in\{1,\ldots,n\}$
we have $c_{ij}>e$, then by (\ref{19}) in Lemma \ref{lemma2} we
have 
\[
c_{ij}=w(a_{i})\div w(a_{j})>e,
\]
which is equivalent to $w(a_{i})>w(a_{j})$ and condition \emph{POP}
is satisfied.
Moreover, by Lemma \ref{lemma2}, it holds that $c_{ij}>c_{kl}$ if and
only if 
\[
w(a_{i})\div w(a_{j})>w(a_{k})\div w(a_{l}),
\]
hence, (\ref{17}) is satisfied. 
\end{proofQED}

\begin{lemma}
\label{lemma3} Let $C=[c_{ij}]$ be an $\odot$-reciprocal PC matrix
and $w=(w_{1},\ldots,w_{n})$ be a priority vector. Then for all $i,j\in\{1,\ldots,n\}$
\begin{equation}
\mathscr{E}(C,w)^{(-1)}\odot w(a_{i})\div w(a_{j})\leq c_{ij}\leq\mathscr{E}(C,w)\odot w(a_{i})\div w(a_{j}).\label{20}
\end{equation}
\end{lemma}

\begin{proofQED}
By (\ref{11}), (\ref{12}) we obtain 
\begin{equation}
\mathscr{E}(C,w)\geq\max\{\epsilon(i,j,w),(\epsilon(i,j,w))^{(-1)}\}\geq\epsilon(i,j,w)=c_{ij}\odot w(a_{j})\div w(a_{i}),\label{21}
\end{equation}
\begin{equation}
\mathscr{E}(C,w)\geq\max\{\epsilon(i,j,w),(\epsilon(i,j,w)^{(-1)}\}\geq\epsilon(i,j,w)^{(-1)}=c_{ji}\odot w(a_{i})\div w(a_{j}),\label{22a}
\end{equation}
hence, when multiplied both sides of (\ref{21}) by $w(a_{i})\div w(a_{j})$,
and both sides of (\ref{22a}) by $w(a_{j})\div w(a_{i})$, we get
\begin{equation}
\mathscr{E}(C,w)\odot w(a_{i})\div w(a_{j})\geq c_{ji}\odot w(a_{i})\div w(a_{j})\odot w(a_{j})\div w(a_{i})=c_{ij}\odot e=c_{ij}.\label{23a}
\end{equation}
\begin{equation}
\mathscr{E}(C,w)^{(-1)}\odot w(a_{i})\div w(a_{j})\leq c_{ij}\odot w(a_{j})\div w(a_{i})\odot w(a_{i})\div w(a_{j})=c_{ij}\odot e=c_{ij}.\label{24a}
\end{equation}
Combining (\ref{23a}) and (\ref{24a}) we obtain (\ref{20}). 
\end{proofQED}

Now, we turn our attention to $\odot$-inconsistent $\odot$-reciprocal \emph{PC}
matrix. The following theorem gives sufficient conditions for validity
of \emph{POP} and \emph{POIP}.

\begin{theorem}\label{theorem2} 
Let $C=[c_{ij}]$ be an $\odot$-reciprocal PC matrix, and let
$w=\left[w(a_{1}),\ldots,w(a_{n})\right]^{T}$ be a priority vector,
let $i,j,k,l\in\{1,\ldots,n\}$.
 If 
\[
c_{ij}>\mathscr{E}(C,w),c_{kl}>\mathscr{E}(C,w)
\]
and 
\begin{equation}
c_{ij}\div c_{kl}>(\mathscr{E}(C,w))^{2},\label{222}
\end{equation}
then 
\[
w(a_{i})>w(a_{j}),w(a_{k})>w(a_{l})
\]
and 
\[
w(a_{i})\div w(a_{j})>w(a_{k})\div w(a_{l}),
\]
i.e., condition POP and also POIP is satisfied. 
\end{theorem}

\begin{proofQED}
If for some $i,j\in\{1,\ldots,n\}$ we have $c_{ij}>\mathscr{E}(C,w)$,
then by (\ref{20}) in Lemma \ref{lemma3} and $c_{ij}>\mathscr{E}(C,w)$
we obtain 
\begin{equation}
\mathscr{E}(C,w)\odot w(a_{i})\div w(a_{j})\geq c_{ij}>\mathscr{E}(C,w),\label{22}
\end{equation}
which implies, when ``multiplied'' both
sides of (\ref{2}) by $\mathscr{E}(C,w)^{(-1)}$, 
\[
w(a_{i})\div w(a_{j})>e,
\]
i.e., $w_{i}>w_{j}$, hence condition POP is satisfied.
 Similarly, if for some $k,l\in\{1,\ldots,n\}$ we have $c_{ij}>\mathscr{E}(C,w),c_{kl}>\mathscr{E}(w)$,
then we obtain $w(a_{i})>w(a_{j})$ and $w(a_{k})>w(a_{l})$, i.e.
POP condition.
Moreover, by (\ref{20}) in Lemma \ref{lemma3} we obtain 
\begin{equation}
\mathscr{E}(C,w)\odot w(a_{i})\div w(a_{j})\geq c_{ij},
\end{equation}
\begin{equation}
\mathscr{E}(C,w)\odot w(a_{l})\div w(a_{k})\geq c_{lk}=c_{kl}^{(-1)}.\label{23}
\end{equation}
By ``$\odot$-multiplying'' left sides and right
sides of (\ref{22}) and (\ref{23}), we obtain 
\begin{equation}
(\mathscr{E}(C,w))^{2}\odot(w(a_{i})\div w(a_{j}))\odot(w(a_{k})\div w(a_{l}))^{(-1)}\geq c_{ij}\div c_{kl}.\label{24}
\end{equation}
If we assume that $(w(a_{i})\div w(a_{j}))\odot(w(a_{k})\div w(a_{l}))^{(-1)}\leq e,$
which is equivalent to $w(a_{i})\div w(a_{j})\leq w(a_{k})\div w(a_{l})$,
then by (\ref{24}) we obtain 
\[
(\mathscr{E}(C,w))^{2}\geq c_{ij}\div c_{kl}.
\]

This result, however, is in contradiction with (\ref{222}), hence, it
must be $w(a_{i})\div w(a_{j})>w(a_{k})\div w(a_{l})$, therefore,
condition \emph{POIP} is satisfied. 
\end{proofQED}

\begin{remark}
Notice that in the previous lemmas and theorems, there is no special
assumption concerning the method for generating the priority vector
$w=\left[w(a_{1}),\ldots,w(a_{n})\right]^{T}$.  The priority vector,
or, vector of weights $w$, may be an arbitrary positive vector with
normalized elements. Specifically, in the case of multiplicative
alo-group of positive real numbers
$\mathcal{R}_{+}=(\mathbb{R}_{+},\cdot,\leq)$ with some field
operation +, we may use 
\emph{EVM}, \emph{GMM} or any other priority vector
generating method. In the following section we shall
investigate a generalized version of \emph{GMM}. 
\end{remark}

\section{Generalized Geometric Mean Method (GGMM)}%
\label{sec:Generalized-Geometric-Mean}

Following the Geometric Mean Method (GMM) we define the \emph{Generalized
Geometric Mean Method (GGMM)}, where the weight of $i$-th alternative
is given by the $\odot$-mean of the $i$-th row of $C=[c_{ij}]$. 
\begin{definition}
\label{def:The generalized inconsistency index} Let $C=[c_{ij}]$
be a reciprocal PC matrix. Let for $i,j,k\in\{1,...,n\}$ 
\begin{equation}
e(i,j,k)\overset{\textit{df}}{=}c_{ij}\odot c_{jk}\odot c_{ki},\label{101}
\end{equation}
and similarly let us define 
\begin{equation}
\eta(i,j,k)\overset{\textit{df}}{=}\max\{e(i,j,k),(e(i,j,k))^{(-1)}\}.\label{102}
\end{equation}
The \emph{generalized inconsistency index} of the PC matrix $C$ is
defined as 
\begin{equation}
\textit{GI}(C)\overset{\textit{df}}{=}\max\{\eta(i,j,k)|i,j,k\in\{1,...,n\}\}.\label{16}
\end{equation}
\end{definition}

\begin{lemma}
\label{lemma4} Let $C=[c_{ij}]$ be a reciprocal PC matrix, $w=(w_{1},\ldots,w_{n})$
be a priority vector defined as 
\begin{equation}
w=\delta \odot \left[\left(\bigodot_{r=1}^{n}c_{1r}\right)^{(\frac{1}{n})},\ldots,\left(\bigodot_{r=1}^{n}c_{nr}\right)^{(\frac{1}{n})}\right]^{T},\label{13}
\end{equation}
where $\delta$ is a scaling factor, $\delta=\left(\bigodot_{r=1}^{n}c_{1r}\right)^{(-\frac{1}{n})}\odot\ldots\odot\left(\bigodot_{r=1}^{n}c_{nr}\right)^{(-\frac{1}{n})}.$
The individual weights are given as 
\begin{equation}
w_{i}=\delta \odot \left(\bigodot_{r=1}^{n}c_{ir}\right)^{(\frac{1}{n})},i\in\{1,...,n\}.\label{14}
\end{equation}
Then the global error index of $C$ is always less or equal to the
generalized inconsistency index, i.e., 
\begin{equation}
\mathscr{E}(C,w)\le\textit{GI}(C).\label{103}
\end{equation}
\end{lemma}

\begin{proofQED}
Providing the use of GGMM we have 
\[
c_{ij}\odot w(a_{j})\div w(a_{i})=c_{ij} \odot \left(\bigodot_{k=1}^{n}c_{jk}\div\bigodot_{k=1}^{n}c_{ik}\right)^{(\frac{1}{n})}
\]

thus, 

\[
c_{ij}\odot w(a_{j})\div w(a_{i})=\left(\bigodot_{k=1}^{n}c_{ij}\odot c_{jk}\div\bigodot_{k=1}^{n}c_{ik}\right)^{(\frac{1}{n})}=\left(\bigodot_{k=1}^{n}c_{ij}\odot c_{jk}\odot c_{ki}\right)^{(\frac{1}{n})}
\]

However, it holds that

\[
\left(\bigodot_{k=1}^{n}c_{ij}\odot c_{jk}\odot c_{ki}\right)^{(\frac{1}{n})}\le\underset{k\in\{1,...,n\}}{\max}\{c_{ij}\odot c_{jk}\odot c_{ki}\}=
\]
\begin{equation}
\underset{k\in\{1,...,n\}}{\max}\{\eta(i,j,k)\},\label{104}
\end{equation}

hence, $\mathscr{E}(C,w)\le\textit{GI}(C).$
\end{proofQED}

Now, we shall derive a relationship between \emph{Koczkodaj's} inconsistency
index (\ref{8}) and the \emph{generalized inconsistency index} (\ref{16})
in the multiplicative alo-group $\mathcal{R}_{+}=(\mathbb{R}_{+},\cdot,\leq)$
together with the additional field operation +. First, we modify Theorem
\ref{theorem2} with respect to the above mentioned \emph{Koczkodaj's}
inconsistency index $\textit{KI}(C)$ and the new generalized inconsistency
index $\textit{GI}(C)$, see \cite[Corollary 1]{Kulakowski2015noop}.
\begin{theorem}
\label{theorem3} Let $C=[c_{ij}]$ be an $\odot$-reciprocal PC matrix,
$w=\left[w(a_{1}),\ldots,w(a_{n})\right]^{T}$ be a priority vector
generated by GGMM, i.e., (\ref{13}) and (\ref{14}).
 If 
\[
c_{ij}>\frac{1}{1-\textit{KI}(C)},c_{kl}>\frac{1}{1-\textit{KI}(C)}
\]
and 
\begin{equation}
c_{ij}\div c_{kl}>\left(\frac{1}{1-\textit{KI}(C)}\right)^{2},\label{2222}
\end{equation}
then 
\[
w(a_{i})>w(a_{j}),w(a_{k})>w(a_{l}),
\]
and 
\[
w(a_{i})\div w(a_{j})>w(a_{k})\div w(a_{l}),
\]
i.e., condition POP and also condition POIP is satisfied. 
\end{theorem}

\begin{proofQED}
Comparing (\ref{8}) and (\ref{12}), (\ref{16}) we easily derive
the relation between $\textit{KI}(C)$ and $\textit{GI}(C)$ as follows
\begin{equation}
\textit{GI}(C)=\frac{1}{1-\textit{KI}(C)}.\label{30}
\end{equation}
If for some $i,j,k,l\in\{1,\ldots,n\}$ we have $c_{ij}>\textit{GI}(C)=\frac{1}{1-\textit{KI}(C)},c_{kl}>\textit{GI}(C)$,
then by \emph{Theorem} \ref{theorem2} and \emph{Lemma} \ref{lemma4}
we obtain $w(a_{i})>w(a_{j})$ and $w(a_{k})>w(a_{l})$. 
\end{proofQED}

\begin{example}
\label{ex:example} We consider an illustrating example of $4\times4$
\emph{PC} matrix $C$ in the usual multiplicative alo-group of positive
real numbers $\mathcal{R}_{+}=(\mathbb{R}_{+},\cdot,\leq)$, as follows:

\[
C=\left[\begin{array}{cccc}
1 & \frac{5}{2} & 3 & 5\\
\frac{2}{5} & 1 & 2 & 4\\
\frac{1}{3} & \frac{1}{2} & 1 & 3\\
\frac{1}{5} & \frac{1}{4} & \frac{1}{3} & 1
\end{array}\right].
\]
The priority vector generated by \emph{GGMM} (in fact \emph{GMM})
is:
$$
w=[0.494,\,0.2675,\,0.168,\,0.072]^{T},
$$
hence, the newly proposed generalized `$\cdot$'-inconsistency index
$\textit{GI}(C)=2.00$, $\textit{GI}(C)^{2}=4.00$ and \emph{Koczkodaj's}
inconsistency index $\textit{KI}(C)=0,5$.  We conclude that if
$c_{ij}>2.00$, then $w(a_{i})>w(a_{j})$. It is clear, that \emph{POP}
condition holds for all elements located above the main diagonal of
\emph{PC} matrix~$C$.  Moreover, $c_{12}=2.50$, $c_{23}=2.00$ and
$c_{12}/c_{23}=1.25.$ We obtain $w_{1}/w_{2}=1.85$ and
$w_{2}/w_{3}=1.59$, hence, $w_{1}/w_{2}>w_{2}/w_{3}.$ Here,
\emph{POIP} condition is satisfied. 
\end{example}

\section{Discussion and summary}

Is it possible to meet the \emph{COP} criteria
\cite{BanaeCosta2008acao} when the ranking method is \emph{EVM}? Would
it be possible to use \emph{GMM} instead of \emph{EVM} while
preserving \emph{COP}?  Theorem \ref{theorem2} provides the evidence
that as long as the result $c_{ij}$ of the direct comparison of the
$i$-th and $j$-th alternative is large enough it is possible. In such
a case the lower limit for $c_{ij}$ is given by the global error index
$\mathscr{E}(C,w)$ defined for any priority vector $w$. Hence, the
minimal value of $c_{ij}$ guaranting that \emph{COP} is met depends on
$w$, regardless of how $w$ is obtained. 

An even more surprising conclusion comes from Theorem~\ref{theorem3}.
Assuming that the weight vector was obtained using \emph{GMM (GGMM)}
the minimal value of $c_{ij}$---which guarantees compliance with the
\emph{COP} criteria---depends only on the inconsistency of the
\emph{PC} matrix. In particular this means that by improving the
consistency among the pairwise comparisons we are able to make the
\emph{PC} matrix comply with \emph{COP}. Interestingly, a similar
situation occurs in the case when the priority deriving method is
\emph{EVM} \cite{Kulakowski2015otpo}.  This raises the question
whether a similar property can be observed for any priority deriving
method.  This question remains unanswered today, however, it seem to
be an interesting direction for further research. 

By abstracting into an alo-group, we define $\textit{GI}$ a new
\emph{generalized inconsistency index} based only on the group
operation $\odot$. We showed the relationship between the triad
$\textit{GI}$, $\textit{KI}$ (\emph{Koczkodaj's} inconsistency index)
and \emph{COP}. The particulars of this relationship have been
examined in the example at the end of
Section~\ref{sec:Generalized-Geometric-Mean}. 

\section*{Acknowledgements }

The research was supported by the National Science Centre, Poland, as
a part of the project no. 2017/25/B/HS4/01617, by the Ko{\'s}ciuszko
Foundation grant exchange with California State University Channel
Islands, and also by project GACR Nr.~18-01246S, Czech Republic.

\section*{References}

\bibliographystyle{plain}
\bibliography{papers_biblio_reviewed}

\begin{thebibliography}{10}

\bibitem{CavalloDApuzzo2009aguf}
L.~D'Apuzzo B.~Cavallo.
\newblock {A general unified framework for pairwise comparison matrices in
  multicriteria methods}.
\newblock {\em International Journal of Intelligent Systems}, 24(4):377--398,
  2009.

\bibitem{BanaeCosta2008acao}
C.~A. Bana~e Costa and J.~Vansnick.
\newblock {A critical analysis of the eigenvalue method used to derive
  priorities in AHP}.
\newblock {\em European Journal of Operational Research}, 187(3):1422--1428,
  June 2008.

\bibitem{Bourbaki1990aii}
N.~Bourbaki.
\newblock {\em {Algebra II}}.
\newblock Springer Verlag, Heidelberg-New York-Berlin, 1990.

\bibitem{Bozoki2010ooco}
S.~Boz{\'o}ki, J.~F{\"u}l{\"o}p, and L.~R{\'o}nyai.
\newblock On optimal completion of incomplete pairwise comparison matrices.
\newblock {\em Mathematical and Computer Modelling}, 52(1--2):318 -- 333, 2010.

\bibitem{Bozoki2008osak}
S.~Boz{\'o}ki and T.~Rapcs{\'a}k.
\newblock {On Saaty's and Koczkodaj's inconsistencies of pairwise comparison
  matrices}.
\newblock {\em Journal of Global Optimization}, 42(2):157--175, 2008.

\bibitem{Brunelli2015itta}
Matteo Brunelli.
\newblock {\em {Introduction to the Analytic Hierarchy Process}}.
\newblock SpringerBriefs in Operations Research. Springer International
  Publishing, Cham, 2015.

\bibitem{Colomer2011rlfa}
J.~M. Colomer.
\newblock {Ramon Llull: from `Ars electionis' to social choice theory}.
\newblock {\em Social Choice and Welfare}, 40(2):317--328, October 2011.

\bibitem{Crawford1987tgmp}
G.~B. Crawford.
\newblock The geometric mean procedure for estimating the scale of a judgement
  matrix.
\newblock {\em Mathematical Modelling}, 9(3--5):327 -- 334, 1987.

\bibitem{Faliszewski2009lacv}
P.~Faliszewski, E.~Hemaspaandra, L.~A. Hemaspaandra, and J.~Rothe.
\newblock {Llull and Copeland Voting Computationally Resist Bribery and
  Constructive Control}.
\newblock {\em J. Artif. Intell. Res. (JAIR)}, 35:275--341, 2009.

\bibitem{Fechner1966eop}
G.~T. Fechner.
\newblock {\em Elements of psychophysics}, volume~1.
\newblock Holt, Rinehart and Winston, New York, 1966.

\bibitem{Fedrizzi2007ipca}
M.~Fedrizzi and S.~Giove.
\newblock {Incomplete pairwise comparison and consistency optimization}.
\newblock {\em European Journal of Operational Research}, 183(1):303--313,
  2007.

\bibitem{Fulop2012oscp}
J.~F{\"u}l{\"o}p, W.~W. Koczkodaj, and S.~J. Szarek.
\newblock On some convexity properties of the least squares method for pairwise
  comparisons matrices without the reciprocity condition.
\newblock {\em J. Global Optimization}, 54(4):689--706, 2012.

\bibitem{Greco2011fk}
S.~Greco, B.~Matarazzo, and R.~S{\l}owi{\'n}ski.
\newblock Dominance-based rough set approach on pairwise comparison tables to
  decision involving multiple decision makers.
\newblock In JingTao Yao, Sheela Ramanna, Guoyin Wang, and Zbigniew Suraj,
  editors, {\em Rough Sets and Knowledge Technology}, volume 6954 of {\em
  Lecture Notes in Computer Science}, pages 126--135. Springer Berlin
  Heidelberg, 2011.

\bibitem{Ho2017tsot}
W.~Ho and X.~Ma.
\newblock {The state-of-the-art integrations and applications of the analytic
  hierarchy process}.
\newblock {\em European Journal of Operational Research}, September 2017.

\bibitem{Ho2008iahp}
William Ho.
\newblock {Integrated analytic hierarchy process and its applications - A
  literature review}.
\newblock {\em European Journal of Operational Research}, 186(1):18--18, March
  2008.

\bibitem{Ishizaka2009ahpa}
A.~Ishizaka and A.~Labib.
\newblock {Analytic hierarchy process and expert choice: Benefits and
  limitations}.
\newblock {\em OR Insight}, 22(4):201--220, 2009.

\bibitem{Ishizaka2011rotm}
A.~Ishizaka and A.~Labib.
\newblock {Review of the main developments in the Analytic Hierarchy Process}.
\newblock {\em Expert Systems with Applications}, 38(11):14336--14345, October
  2011.

\bibitem{Janicki2011ropc}
R.~Janicki and Y.~Zhai.
\newblock {Remarks on Pairwise Comparison Numerical and Non-numerical
  Rankings}.
\newblock In {\em Rough Sets and Knowledge Technology}, pages 290--300.
  Springer, Berlin, Heidelberg, Berlin, Heidelberg, October 2011.

\bibitem{Janicki2012oapc}
R.~Janicki and Y.~Zhai.
\newblock On a pairwise comparison-based consistent non-numerical ranking.
\newblock {\em Logic Journal of the IGPL}, 20(4):667--676, 2012.

\bibitem{Koczkodaj1993ando}
W.~W. Koczkodaj.
\newblock A new definition of consistency of pairwise comparisons.
\newblock {\em Math. Comput. Model.}, 18(7):79--84, October 1993.

\bibitem{Koczkodaj2014oaoi}
W.W. Koczkodaj and R.~Szwarc.
\newblock On axiomatization of inconsistency indicators for pairwise
  comparisons.
\newblock {\em Fundamenta Informaticae}, 4(132):485--500, 2014.

\bibitem{Kou2016pcmi}
G.~Kou, D.~Ergu, C.~S. Lin, and Y.~Chen.
\newblock {Pairwise comparison matrix in multiple criteria decision making}.
\newblock {\em Technological and Economic Development of Economy},
  22(5):738--765, 2016.

\bibitem{Kulakowski2013hrea}
K.~Ku{\l}akowski.
\newblock {Heuristic Rating Estimation Approach to The Pairwise Comparisons
  Method}.
\newblock {\em Fundamenta Informaticae}, 133:367--386, 2014.

\bibitem{Kulakowski2015noop}
K.~Ku{\l}akowski.
\newblock {Notes on Order Preservation and Consistency in AHP}.
\newblock {\em European Journal of Operational Research}, 245(1):333--337,
  2015.

\bibitem{Kulakowski2015otpo}
K.~Ku{\l}akowski.
\newblock On the properties of the priority deriving procedure in the pairwise
  comparisons method.
\newblock {\em Fundamenta Informaticae}, 139(4):403 -- 419, July 2015.

\bibitem{kulakowski2014hreg}
K.~Ku{\l}akowski, K.~Grobler-D{\k e}bska, and J.~W{\k a}s.
\newblock Heuristic rating estimation: geometric approach.
\newblock {\em Journal of Global Optimization}, 62, 2014.

\bibitem{Liberatore2008tahp}
M.~J. Liberatore and R.~L. Nydick.
\newblock {The analytic hierarchy process in medical and health care decision
  making: A literature review}.
\newblock {\em European Journal of Operational Research}, 189(1):14--14, August
  2008.

\bibitem{Mikhailov2003dpff}
L.~Mikhailov.
\newblock {Deriving priorities from fuzzy pairwise comparison judgements}.
\newblock {\em Fuzzy Sets and Systems}, 134(3):365--385, March 2003.

\bibitem{Ramik2015ibfp}
J.~Ramik.
\newblock {Isomorphisms between fuzzy pairwise comparison matrices}.
\newblock {\em Fuzzy Optimization and Decision Making}, 14:199--209, 2015.

\bibitem{Saaty1977asmf}
T.~L. Saaty.
\newblock A scaling method for priorities in hierarchical structures.
\newblock {\em Journal of Mathematical Psychology}, 15(3):234 -- 281, 1977.

\bibitem{Saaty2005taha}
T.~L. Saaty.
\newblock The analytic hierarchy and analytic network processes for the
  measurement of intangible criteria and for decision-making.
\newblock In {\em Multiple Criteria Decision Analysis: State of the Art
  Surveys}, volume~78 of {\em International Series in Operations Research and
  Management Science}, pages 345--405. Springer New York, 2005.

\bibitem{Saaty2013dk}
T.~L. Saaty.
\newblock {On the Measurement of Intangibles. A Principal Eigenvector Approach
  to Relative Measurement Derived from Paired Comparisons}.
\newblock {\em Notices of the American Mathematical Society}, 60(02):192--208,
  February 2013.

\bibitem{soltys-fund-2015}
Barbara Sandrasagra and Michael Soltys.
\newblock Complex ranking procedures.
\newblock {\em Fundamenta Informaticae Special Issue on Pairwise Comparisons},
  144(3-4):223--240, 2016.

\bibitem{Smith2004aada}
J.~E. Smith and D.~Von~Winterfeldt.
\newblock {Anniversary article: decision analysis in management science}.
\newblock {\em Management Science}, 50(5):561--574, 2004.

\bibitem{Subramanian2012aroa}
N.~Subramanian and R.~Ramanathan.
\newblock {A review of applications of Analytic Hierarchy Process in operations
  management}.
\newblock {\em International Journal of Production Economics}, 138(2):215--241,
  August 2012.

\bibitem{Thurstone27aloc}
L.~L. Thurstone.
\newblock A law of comparative judgment, reprint of an original work published
  in 1927.
\newblock {\em Psychological Review}, 101:266--270, 1994.

\bibitem{Vaidya2006ahpa}
O.~S. Vaidya and S.~Kumar.
\newblock {Analytic hierarchy process: An overview of applications}.
\newblock {\em European Journal of Operational Research}, 169(1):1--29,
  February 2006.

\bibitem{Yuen2013fcnp}
K.~K.~F. Yuen.
\newblock Fuzzy cognitive network process: Comparison with fuzzy analytic
  hierarchy process in new product development strategy.
\newblock {\em Fuzzy Systems, IEEE Transactions on}, PP(99):1--1, 2013.

\end{thebibliography}

\end{document}